\documentclass[11pt]{article}
\usepackage{moriond,epsfig}
\def\jpsi{J/$\psi$}

\begin{document}
\vspace*{4cm}
\title{Dimuon production in p-nucleus and nucleus-nucleus collisions: the NA60 experiment}
\author{Roberta Arnaldi for the NA60 Collaboration:\\
  R.~Arnaldi\,$^1$,
  R.~Averbeck\,$^2$
  K.~Banicz\,$^3$,
  K.~Borer\,$^4$,
  J.~Buytaert\,$^5$,
  J.~Castor\,$^6$,
  B.~Chaurand\,$^7$,
  W.~Chen\,$^8$,
  B.~Cheynis\,$^9$,
  C.~Cical\'o\,$^{10}$,
  A.~Colla\,$^1$,
  P.~Cortese\,$^1$,
  S.~Damjanovic\,$^3$,
  A.~David\,$^{11,5}$,
  A.~de~Falco\,$^{10}$,
  N.~de~Marco\,$^1$,
  A.~Devaux\,$^6$,
  A.~Drees\,$^2$,
  L.~Ducroux\,$^9$,
  H.~En'yo\,$^{12}$,
  A.~Ferretti\,$^1$,
  M.~Floris\,$^{10}$,
  P.~Force\,$^6$,
  A.~Grigorian\,$^{13}$,
  J.-Y.~Grossiord\,$^9$,
  N.~Guettet\,$^{5,6}$,
  A.~Guichard\,$^9$,
  H.~Gulkanian\,$^{13}$,
  J.~Heuser\,$^{12}$,
  M.~Keil\,$^5$,
  L.~Kluberg\,$^7$,
  Z.~Li\,$^8$,
  C.~Louren\c{c}o\,$^5$,
  J.~Lozano\,$^{11}$,
  F.~Manso\,$^6$,
  P.~Martins\,$^{11}$,
  A.~Masoni\,$^{10}$,
  A.~Neves\,$^{11}$,   
  H.~Ohnishi\,$^{12}$,
  C.~Oppedisano\,$^1$,
  P.~Parracho\,$^{5,11}$,
  G.~Puddu\,$^{10}$,
  E.~Radermacher\,$^5$,
  P.~Ramalhete\,$^{11}$,
  P.~Rosinsk\'y\,$^5$,
  E.~Scomparin\,$^1$,
  J.~Seixas\,$^{11}$,
  S.~Serci\,$^{10}$,
  R.~Shahoyan\,$^{11}$,
  P.~Sonderegger\,$^{11}$,
  H.J.~Specht\,$^3$,
  R.~Tieulent\,$^9$,
  G.~Usai\,$^{10}$,
  H.~Vardanyan\,$^{13}$,
  R.~Veenhof\,$^{11}$ and
  H.~W\"ohri\,$^5$}

\address{
$^1$ Universit\`a di Torino and INFN, Turin, Italy;
$^2$ SUNY at Stony Brook, New York, USA;\\
$^3$ Heidelberg University, Heidelberg, Germany;
$^4$ LHEP, University of Bern, Bern, Switzerland;\\
$^5$ CERN, Geneva, Switzerland;
$^6$ LPC, Univ.\ Blaise Pascal and CNRS-IN2P3, Clermont-Ferrand, France;
$^7$ LLR, Ecole Polytechnique and CNRS-IN2P3, Palaiseau, France;\\
$^8$ BNL, Upton, New York, USA;
$^9$ IPN, Univ.\ Claude Bernard Lyon-I and CNRS-IN2P3, Lyon, France;\\
$^{10}$ Universit\`a di Cagliari and INFN, Cagliari, Italy;
$^{11}$ Instituto Superior T\'ecnico, Lisbon, Portugal;\\
$^{12}$ RIKEN, Wako, Saitama, Japan;
$^{13}$ YerPhI, Yerevan, Armenia}

\maketitle

\abstracts{The measurements of dilepton production in heavy-ion
collisions performed at the CERN SPS provided some of the most
interesting observations done so far in the search for quark gluon
plasma formation.  However, certain aspects in the interpretation of
these measurements remain unclear and require further studies.  Thanks
to a radiation tolerant silicon pixel telescope, NA60 has measured, in
2003, dimuon production in nuclear collisions with unprecedented
accuracy.  The physics topics under study include the J/$\psi$
suppression pattern, the excess of the intermediate mass dimuons and
of low mass dileptons.  In this paper, after briefly summarizing the
NA60 physics motivations and the detector concept, we present some
preliminary results from Indium-Indium collisions at 158~GeV.}

Lattice QCD predicts that, above a critical temperature or energy
density, strongly interacting matter undergoes a phase transition from
hadronic matter to a new state named Quark Gluon Plasma (QGP). In this
new state quarks and gluons are no longer confined into hadrons and
chiral symmetry is restored.  Since 1986, at the CERN SPS, several
experiments searched for this phase transition in heavy ion
collisions.  The results of this research program provided compelling
evidence for the production of a new state of matter in high-energy
Pb-Pb collisions.  However, further work is needed to clarify some
aspects of these ``anomalous'' observations.

The dimuon mass spectrum between the $\phi$ and the J/$\psi$
resonances is dominated by Drell-Yan and simultaneous semi-leptonic
decays of D mesons. The superposition of these two sources describes
the measurements done in p-A collisions, while in A-A collisions the
dimuon mass spectrum shows an excess which grows with the number of
nucleons participating in the interaction.  Two interpretations of
this excess have been considered: it can be due to an unexpected
enhancement of charm production or to thermal dimuons emitted from the
QGP phase. This question will be answered by extrapolating the muon
tracks back to the target: thermal dimuons are emitted directly from
the interaction point, while muons from charmed mesons come from
vertices displaced with respect to the primary one.  Thanks to the
vertex pixel telescope, NA60 can measure the transverse coordinate of
the tracks at the target, with 50~$\mu$m accuracy.

The CERES experiment at the CERN SPS measured the dielectron invariant
mass spectrum in Pb-Au collisions at 158 GeV. A comparison with the
expected sources (mainly light meson decays, which describe the proton
data) shows an excess in the mass range 0.2--0.7~GeV/$c^2$. This
observation has been interpreted as an indication of changes in the
mass and decay width of the $\rho$ meson, maybe due to partial
restoration of the chiral symmetry. However, this result suffers from
lack of statistics and a poor signal-to-background ratio.  To clarify
these issues NA60 will increase the statistics and improve the mass
resolution and signal to background ratio. The excess will also be
studied as a function of the transverse momentum and the collision
centrality.

The J/$\psi$ suppression, as a signature of the formation of a
deconfined state, has been studied by NA38 and NA50.  In Pb-Pb
collisions the J/$\psi$ production pattern, as a function of the
collision centrality, shows that above a certain centrality threshold
the J/$\psi$ yield is considerably lower than expected from the
``nuclear absorption'' curve, derived from proton-nucleus and light
ions data.  One of the current interpretations of this result is that
the dense and hot medium formed in the collisions dissolves the
$\chi_{c}$ resonance, leading to the disappearance of the fraction
($\sim$\,30\,\%) of J/$\psi$ mesons that would otherwise originate
from $\chi_{c}$ decays.  Several questions remain open. What is the
physical variable driving the J/$\psi$ suppression? Is it the number
of participant nucleons? Or the local energy density? Or the average
length of nuclear matter, $L$, traversed by the charmonium state?  The
results from the In-In data will allow NA60 to answer these questions.
Furthermore, in 2004 NA60 will study the A dependence of $\chi_{c}$
production, with a proton beam at 400~GeV, in order to understand the
level of this feed-down source.

In terms of detector layout, the NA60 experiment complements the Muon
Spectrometer and Zero Degree Calorimeter inherited from the NA50
experiment with a completely new target region, where silicon
micro-strip and pixel detectors, integrated in a 2.5~T dipole magnet,
allow us to track the charged particles produced in the collisions and
accurately determine primary and secondary vertices.  By matching the
muons measured in the muon spectrometer to tracks in the silicon
vertex telescope, simultaneously using information on coordinates and
momentum, we are able to overcome the uncertainties introduced by
multiple scattering and energy loss fluctuations, induced by the
crossing of the hadron absorber.  Besides the obvious improvement in
dimuon mass resolution, we are also able to accurately determine the
origin of the muons, and separate prompt dimuons (Drell-Yan, thermal)
from the muon pairs due to decays of  D mesons.

\begin{figure}[ht]
\centering
\begin{tabular}{ccc}
\resizebox{0.32\textwidth}{!}{%
\includegraphics*[bb= 0 1 515 465]{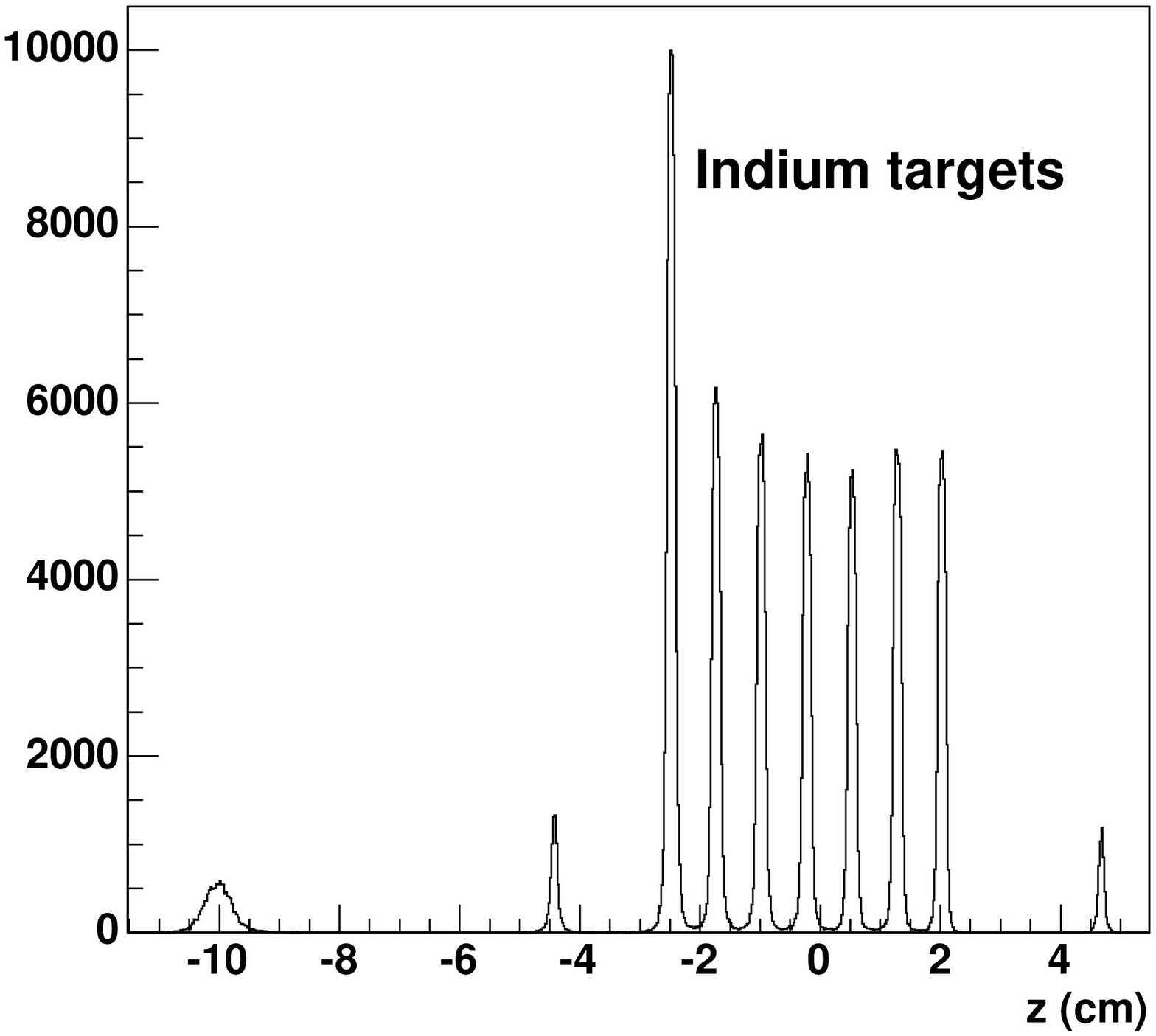}} &
\resizebox{0.305\textwidth}{!}{%
\includegraphics*[bb= 12 160 515 638]{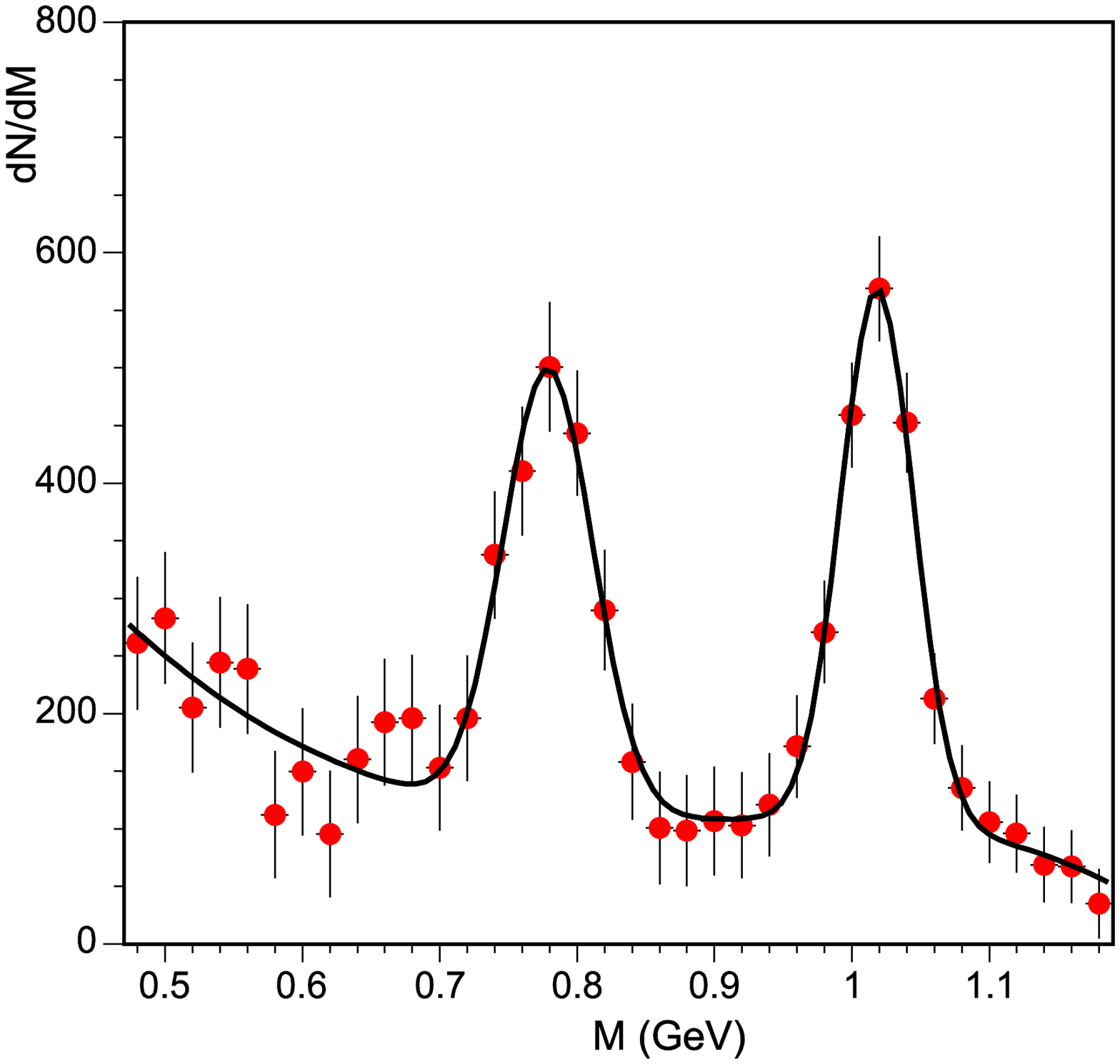}} &
\resizebox{0.305\textwidth}{!}{%
\includegraphics*[bb= 22 9 553 514]{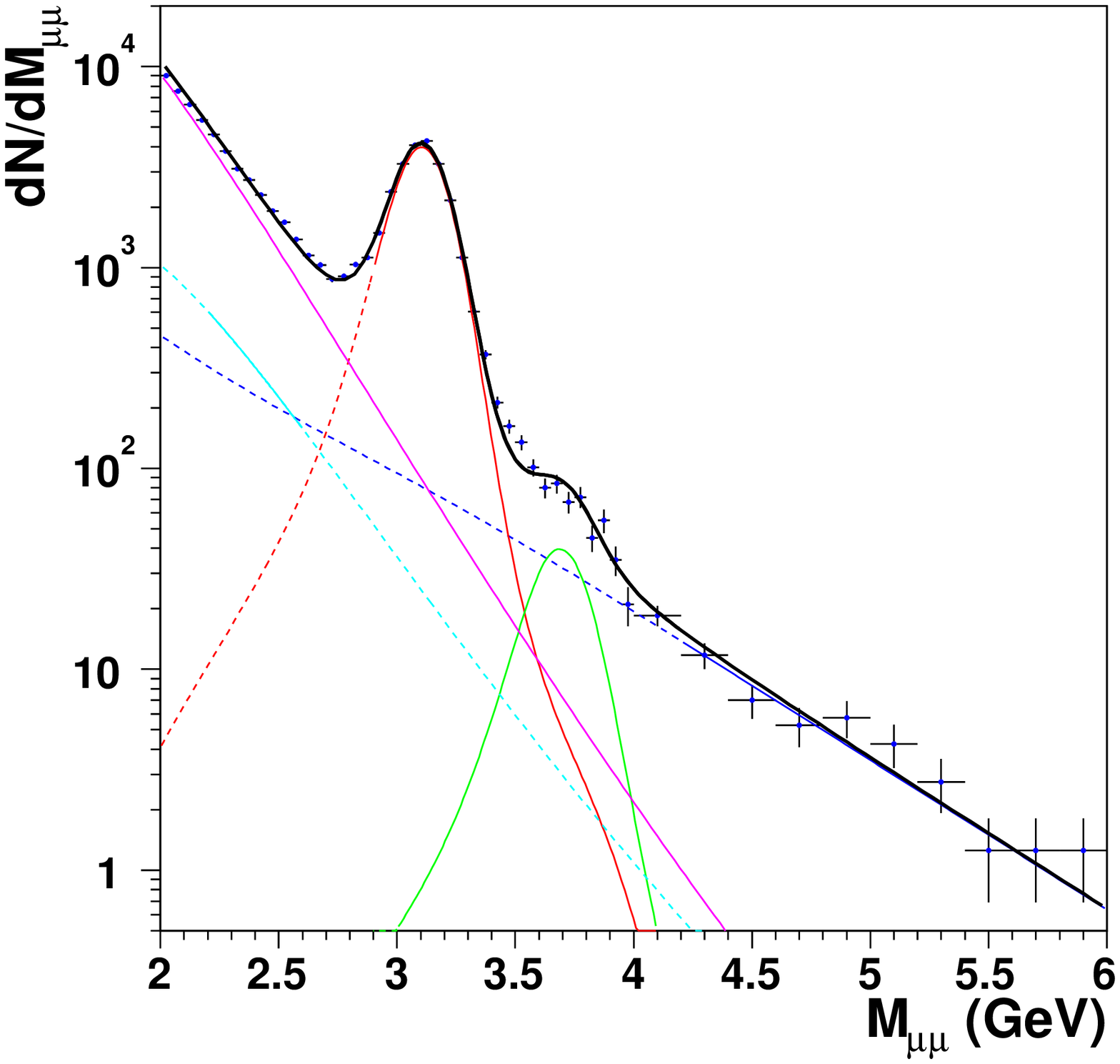}} \\
\end{tabular}
\caption{Distributions measured in Indium-Indium collisions: the
  z-coordinate of the reconstructed vertices (left), the low mass
  dimuon signal after track matching (centre), and the high mass
  spectrum before track matching and background subtraction (right).}
\label{fig1}
\end{figure}

The NA60 detector concept became feasible thanks to the recent
availability of radiation tolerant silicon pixel read-out chips,
developed in view of the LHC experiments.  In the 5-week long Indium
run of October-November 2003, in particular, we could reconstruct the
charged particles trajectories thanks to 11 tracking points provided
by 16 modules covering the muon spectrometer's angular acceptance
($3<\eta_{\rm lab}<4$).  The basic units of these detector planes are
the ALICE1LHCb pixel chips, of $32\times256$ cells of
$425\times50$~$\mu$m$^2$ area.  The kind of accuracy reached with
these detectors can be appreciated by looking at the Z-vertex
distribution (left panel of Fig.~\ref{fig1}) of the interaction
vertices, where we can easily distinguish, with $\sim$\,200~$\mu$m
resolution, the seven Indium targets in-between the two target box
windows (the target was in vacuum).  We can also see the sensor
of a beam tracker module, 10~cm up-stream of the target centre.

Out of the more than 200 million dimuon triggers collected, we expect
to retain around one million low mass signal dimuons, after track
matching and background subtraction.  The second panel of
Fig.~\ref{fig1} shows the extracted signal dimuon spectrum in the
region of the $\omega$ and $\phi$ resonances, on the basis of around
1\,\% of the collected statistics.  It already shows a mass resolution
around 20--25~MeV, in spite of the fact that the reconstruction was
done with a preliminary version of the offline software.  The
combinatorial background resulting from $\pi$ and K decays was
estimated through a mixed-event technique, based on the like-sign muon
pairs.  The analysis of low mass dimuon production can be done down to
very low dimuon transverse momentum and as a function of the collision
centrality.

The muon track matching is not so crucial for the analysis of the
J/$\psi$ meson, clearly visible over the underlying continuum even
with a mass resolution of 100~MeV, as can be seen in the right panel
of Fig.~\ref{fig1}.  Also this figure is made with only a fraction
(around 50\,\%) of the collected data and having applied a severe
event selection procedure, to eliminate pile-up and events with
interactions out of the targets.

The dimuon mass region above 2~GeV contains the J/$\psi$ and
$\psi^\prime$ resonances sitting on a continuum composed of Drell-Yan
dimuons and of muon pairs from decays of D mesons, besides the
combinatorial background from $\pi$ and K decays, which we determine
from the measured like-sign muon pairs.  In order to extract the ratio
between the J/$\psi$ and the Drell-Yan production cross-sections,
integrated over all collision centralities where $E_{\rm
ZDC}<15$~TeV, we fit the opposite-sign dimuon mass distribution to a
superposition of the contributions just mentioned.  The corresponding
mass distributions are evaluated through a detailed Monte Carlo
simulation, using Pythia for the event generation and GEANT for the
reproduction of the detector effects.  The events are then
reconstructed as the real data.  Besides the simulated invariant mass
spectra, these calculations also provide the J/$\psi$ and Drell-Yan
acceptances, $A_{\psi}=12.4$\,\% and $A_{DY(2.9-4.5)}=13.4$\,\%, in
our phase space window: $0<y_{\rm cms}<1$ and $|\cos(\theta_{\rm
CS})<0.5|$.  The fit proceeds in three consecutive steps: first we
determine the Drell-Yan yield from the mass region above 4.2~GeV; then
we determine the charm normalization from the mass window
$2.2<M<2.5$~GeV (having fixed the Drell-Yan from the previous step);
finally we extract the J/$\psi$ and $\psi^\prime$ yields, and the
exact position and width of the J/$\psi$ peak, from the mass region
$2.9<M<4.2$~GeV.  We should note that the ratio $\psi$/DY is
completely insensitive to the specific level of open charm decays we
use and it would change by less than 3\,\% if the background would be
10\,\% higher.

In order to compare with previous results, published by the NA38 and
NA50 experiments, obtained in p-nucleus, S-U, and Pb-Pb collisions, we
refer our Indium-Indium result to the Drell-Yan production
cross-section integrated in the mass domain between 2.9 and 4.5~GeV.
Our analysis gives a $\psi$/DY ratio of $19.5\pm1.6$.  A more detailed
work is on-going but we do not expect that this preliminary value will
change by more than 10\,\%.

\begin{figure}[ht]
\centering
\begin{tabular}{cc}
\resizebox{0.43\textwidth}{!}{%
\includegraphics*[bb= 19 156 549 676]{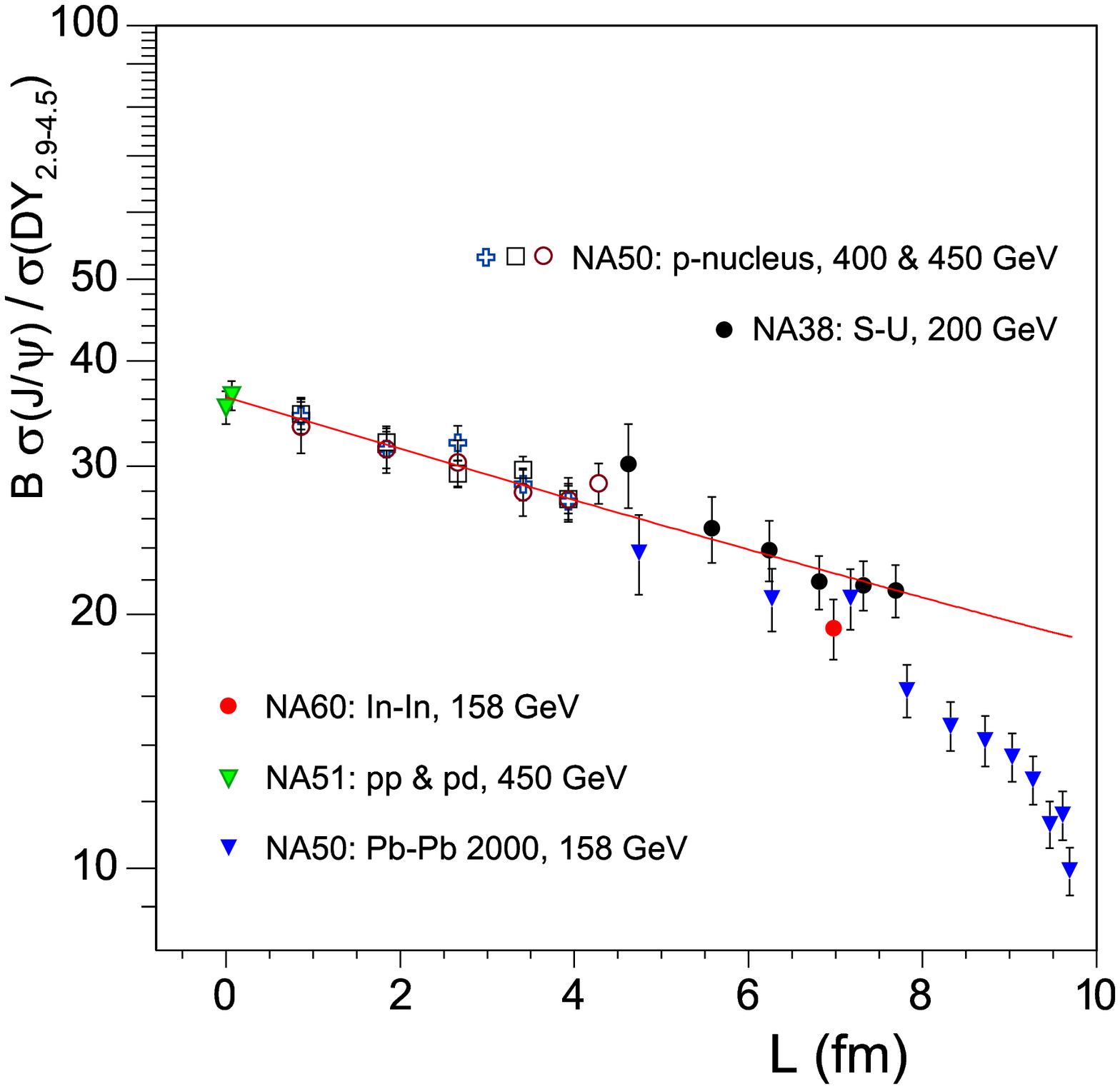}} &
\resizebox{0.43\textwidth}{!}{%
\includegraphics*[bb= 41 156 576 682]{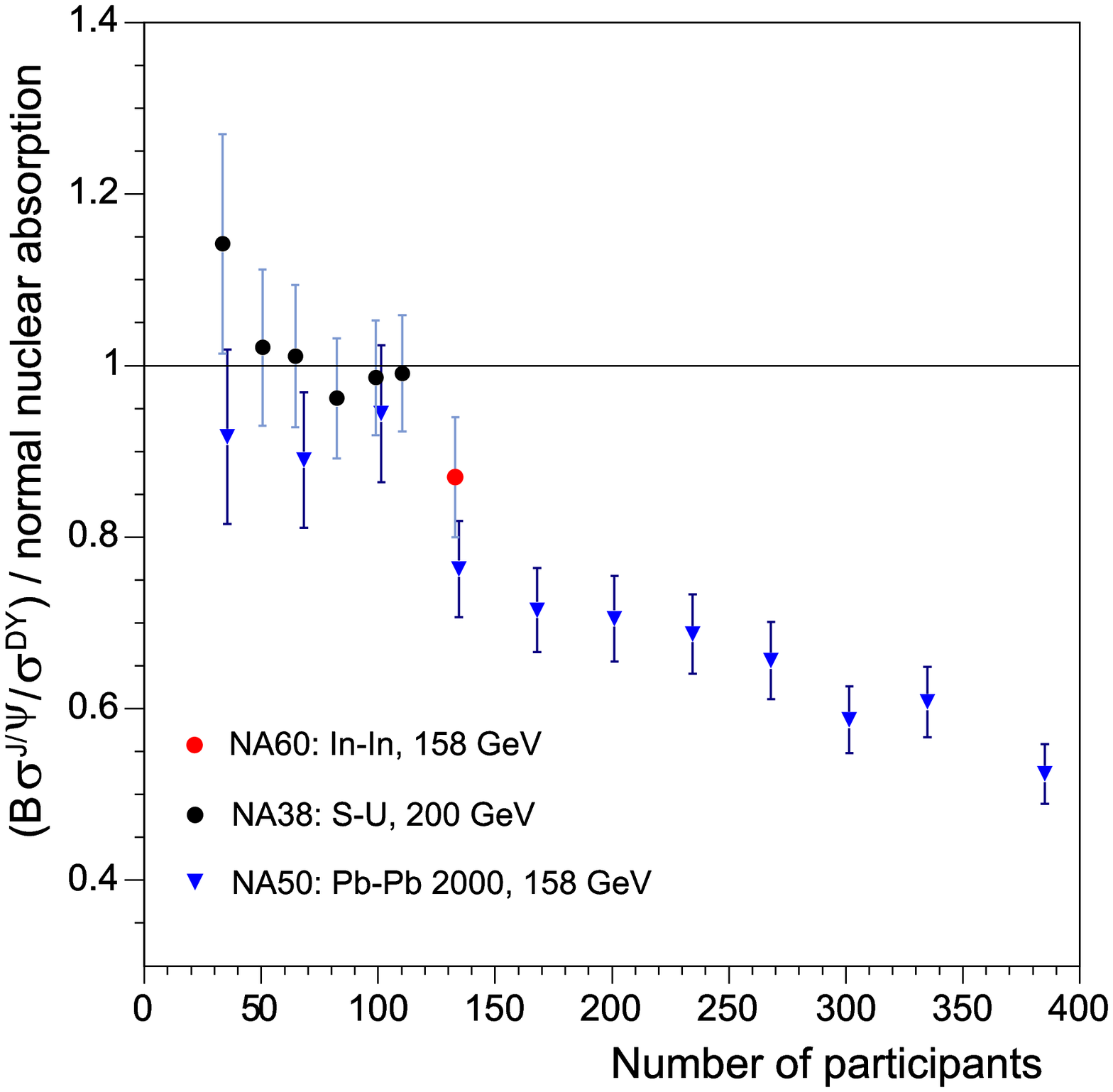}} \\
\end{tabular}
\caption{\jpsi\ suppression pattern versus $L$ (left) and $N_{\rm
    part}$ (right), including the Indium-Indium measurement.}
\label{fig2}
\end{figure}

Figure~\ref{fig2} shows how the Indium-Indium point compares to the
previously established \jpsi\ suppression pattern, both as a function
of $L$, the thickness of nuclear matter traversed by the charmonium
state, on the left panel, and as a function of the number of nucleons
involved in the collision, $N_{\rm part}$, on the right panel.  In the
latter figure the data points are plotted with respect to the normal
nuclear absorption curve, determined from the p-A measurements, which
can be seen in the left panel (continuous line).  Our Indium-Indium
measurement, when referred to the absorption curve, gives the value
$0.87\pm0.07$.  While in the $L$ representation our In-In point sits
(at 7.0~fm) to the left of the most central S-U value, in the $N_{\rm
part}$ plot we see the contrary.  Once all the collected Indium data
will be fully analyzed, we should be able to probe the \jpsi\
suppression pattern as a function of centrality, in the ranges
5.5--7.8 fm in $L$ and 50--200 in $N_{\rm part}$.  By comparing the In-In
and Pb-Pb patterns as a function of different centrality variables, we
should understand which is the variable and, therefore, the physics
mechanism, driving the J/$\psi$ suppression.  In particular, we should
be able to distinguish between the thermal (QGP) and geometrical
(percolation) phase transitions, both resulting in the suppression of
\jpsi\ production but as a function of different variables and with
different thresholds in collision centrality.  It is very important
that the theorists provide now, \emph{before} the centrality
dependence of our Indium data is presented, the corresponding
predictions of their models.

\bigskip

In summary, we have presented preliminary results from the Indium data
collected by NA60 at the end of 2003, revealing a very substantial
improvement in the quality of the measurements with respect to
previous experiments, with important impacts in our understanding of
the low mass dilepton production and of the J/$\psi$ suppression
pattern.  Integrating the Indium data over collision centralities up
to $E_{\rm ZDC}=15$~TeV, we obtain a preliminary value for the ratio
$\psi$/DY, $19.5\pm1.6$, which we compared with previous measurements.
More detailed results, also on open charm production, will be
available soon.  In 2004, high statistics proton-nucleus data will be
taken, providing an accurate reference baseline to understand the
heavy-ion results.

\end{document}